\documentclass[prb,twocolumn,preprintnumbers,amsmath,amssymb,showpacs]{revtex4-1}
\usepackage{graphicx}
\usepackage{amssymb}
\usepackage{mathrsfs}
\usepackage{amsmath}
\usepackage{color}
\usepackage{bbold}
\usepackage{hyperref}
\usepackage{esint}

\newcommand*\D{\mathop{}\!\mathrm{d}}

\newcommand{\beq}{\begin{equation}}
\newcommand{\eeq}{\end{equation}}

\usepackage[capitalise]{cleveref}

\begin{document}

\title{Weyl nodal surfaces}

\author{O\u{g}uz T\"{u}rker}
\email{oguz.tuerker@tu-dresden.de}
\author{Sergej Moroz}
\email{sergej.moroz@tum.de}
\affiliation{ Department of Physics, Technical University of Munich, 85748 Garching, Germany}

\begin{abstract}
We consider three-dimensional fermionic band theories that exhibit Weyl nodal surfaces defined as two-band degeneracies that form closed surfaces in the Brillouin zone. We demonstrate that topology ensures robustness of these objects under small perturbations of a Hamiltonian. This topological robustness is illustrated in several four-band models that exhibit nodal surfaces protected by unitary or anti-unitary symmetries.  Surface states and Nielsen-Ninomiya doubling of nodal surfaces are also investigated.
\end{abstract}


\maketitle


\section{Introduction} \label{intro}

The advent of topological insulators in the last decade deepened  our understanding of interplay of topology and symmetries in band insulators \cite{Hasan2010, Qi2011}. This work culminated in the development of the ten-fold way classification of non-interacting gapped topological phases \cite{Schnyder2008} and the emergence of new symmetry protected topological phases of matter.

In last years the main interest in the field shifted towards systems with band degeneracies \cite{Turner2013, Burkov2015, Armitage2017}. In three dimensions the simplest and most well-studied are Weyl (semi)metals which are distinguished by isolated point-like two-band degeneracies in the Brillouin zone (BZ). Although in condensed matter physics Weyl points appeared first long time ago in the superfluid A phase of $^3$He \cite{volovik1992exotic, volovikbook}, only recently Weyl (semi)metals were discovered experimentally \cite{Xu2015, Lv2015}. As long as inversion or time-reversal symmetry is broken, Weyl points appear generically. They are topological defects and cannot be gapped out individually but must be destroyed only via pair-wise annihilation \cite{Nielsen1981}. Weyl (semi)metals exhibit robust phenomena such as chiral anomaly \cite{Son2013}, anomalous Hall effect \cite{Yang2011} and zero-energy Fermi arc surface states \cite{Wan2011}. More recently Weyl loop (semi)metals \cite{Burkov2011}, where two-band degeneracies take place on closed one-dimensional manifolds also attracted considerable attention. In contrast to Weyl points, nodal loops are not generic, but require some symmetry (such as chiral sub-lattice symmetry) to protect them. The defining feature of a Weyl loop is a non-trivial $\pi$ Berry phase along any closed contour that links with it. In the semimetal regime, any boundary surface, where the loop projects non-trivially, supports drumhead states \cite{Burkov2011, Kim2015}.

\begin{figure}[ht]
\begin{center}
\includegraphics[width=0.5\textwidth]{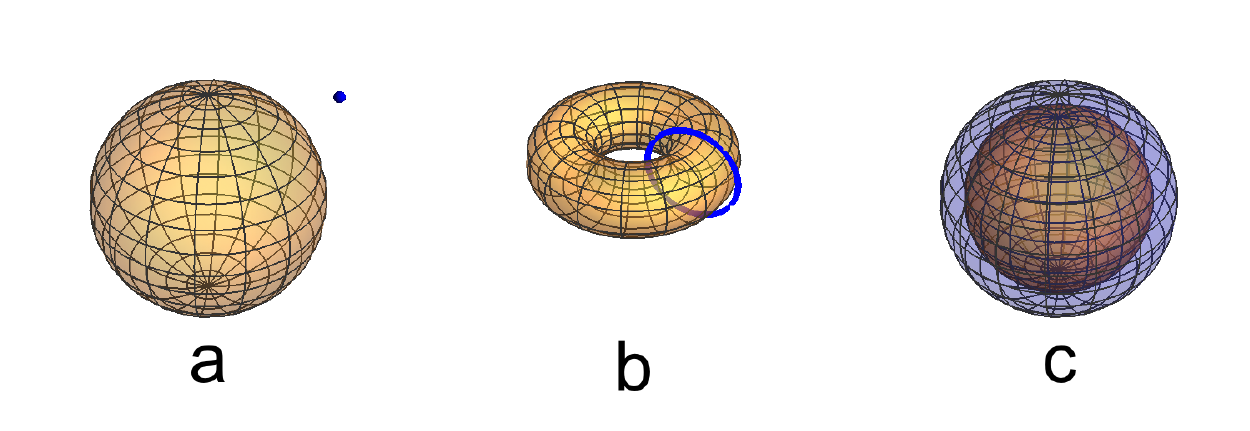}
\caption{Examples of (a) zero-, (b) one- and (c) two-dimensional enclosing manifolds (blue) around nodal surfaces (orange). Although technically a zero-dimensional manifold does not enclose the nodal surface with slight abuse of notation we still refer to it here as an enclosing manifold.}\label{fig1}
\end{center}
\end{figure}

One can make a step further and consider three-dimensional translation-invariant fermionic systems with Weyl nodal surfaces, where two bands touch each other on two-dimensional surfaces in the BZ.   Recently nodal surfaces were predicted to appear in quasi-one-dimensional crystals \cite{QiFeng2016},  graphene networks \cite{Zhong2016},   multi-band superconductors with broken time-reversal symmetry \cite{Agterberg2017, Timm2017} and also were found within the ten-fold way classification of gapless inversion-enriched systems \cite{Bzdusek2017}.
Given a model with a nodal surface it is natural to ask if the nodal structure is robust under certain class of small perturbations of the Hamiltonian. Generically a perturbation can (i) open a gap everywhere and fully destroy the nodal object (ii) gap it out partially leaving behind nodal loops and/or points \footnote{One can imagine a scenario where a closed nodal surface is partially gapped to a nodal surface with boundaries. This case is not discussed in this paper.}, (iii) preserve the nodal surface and not open a gap anywhere on it. As we show in this paper the degree of robustness (i)-(iii) is determined by topology of the perturbed system. To see how it works, consider first a system which has only one nodal surface (tuned for convenience to the Fermi level), but no other gapless objects at the Fermi level such as additional Fermi surfaces, nodal loops or points. Now we enclose the nodal surface of the original model by a lower-dimensional ($d_\text{m}<3$) manifold in the BZ (see Fig. \ref{fig1}). 
 By construction, all bands are gapped on the enclosing manifold. Imagine now that in the original model we can define a topological invariant (such as Chern number, winding number, $\mathbb{Z}_2$ invariant, etc) on this enclosing manifold. In this paper such invariant will be denoted as $c_{d_{\text{m}}}$, where the subscript specifies the dimension of the enclosing manifold.  Importantly, the invariant does not change under a certain class of perturbation of the Hamiltonian and as we show now this has implications for the degree of robustness of the nodal surface under this class of perturbations. Previously, this set of ideas was introduced for determining the robustness of Weyl nodal loops \cite{Fang2015, Fang2016}.
 
 First, we discuss topological invariants defined on a point ($d_\text{m}=0$, Fig. \ref{fig1} a) enclosing manifold. As long as such invariants are different for point manifolds placed inside and outside of the nodal surface in the BZ, the nodal surface cannot be gapped. The reason for that is the following: 
notice that on any imaginary one-dimensional trajectory in the BZ that connects the inner and outer zero-dimensional manifolds  there should be a point, where the energy gap closes allowing the topological invariant to change. In turn this ensures that no gap can open at any point of the nodal surface.

 Second, consider non-trivial topological invariants defined on enclosing manifolds of dimension $d_\text{m}>0$ (Fig. \ref{fig1} b and c). These invariants do not fully protect the nodal surface, but guarantee that it cannot be fully gapped out by perturbations. Generically,   a nodal line (nodal points) should survive in the perturbed system if $d_\text{m}=1$ ($d_\text{m}=2$). We can explain it by following similar arguments as before: in the unperturbed model compute the topological invariant on two enclosing manifolds, one being inside and another outside the nodal surface. Since the topological invariant on the interior manifold  is necessarily trivial (the manifold can be shrunk to a point without encountering a band), a non-trivial invariant on the exterior manifold necessarily implies a non-trivial difference of the interior and exterior invariants which in turn guarantees a lower-dimensional nodal object in between the two enclosing manifolds in the perturbed system. 
 
We emphasize that arguments presented above also imply robustness of general nodal surfaces, not necessarily tuned to the Fermi level. In a general case it is useful to formulate the above arguments in energy-momentum space, as detailed in Appendix \ref{App1}. Our analysis thus does not reduce to previous studies of  robustness of Fermi surfaces \cite{volovikbook, horava2005, Zhao2013, Matsuura2013}.

To apply these ideas in practice, in this paper we investigate nodal surfaces protected by different mechanisms: (i) a nodal surface can be protected by a global internal symmetry. In this case any two bands that carry different quantum numbers with respect to the symmetry generically intersect on a two-dimensional surface in momentum space. Since in the presence of the symmetry the two bands cannot be hybridized, the nodal surface is protected against any small symmetry-preserving perturbations. (ii) in spirit of ten-fold way, Weyl nodal surfaces can be protected by anti-unitary symmetries. These nodal surfaces have already appeared in the literature \cite{Agterberg2017, Bzdusek2017, Timm2017} and here we investigate their robustness. 
We analyze simple models that exhibit nodal surfaces protected by both mechanisms mentioned above.
The leitmotif of our construction is the following:  we start from double-degenerate nodal points or nodal lines and split them in energy.
The resulting continuum four-band models together with minimal lattice extensions thereof are used to analyze the physics of nodal surfaces and investigate their robustness. In particular, we identify symmetries that protect the nodal surface and construct appropriate topological invariants. In addition, we investigate Nielsen-Ninomiya doubling of nodal surfaces in the BZ and look for surface states and topological invariants that protect these states. 
%

\section{Nodal surfaces protected by unitary $U(1)$ symmetry}  \label{nodalU1}
In this section we construct and investigate non-interacting fermionic four-band models that exhibit nodal surfaces protected by an internal unitary $U(1)$ symmetry.
In examples discussed here this symmetry is generated by the axial $\gamma^5$ matrix. More physical realization of a global internal $U(1)$ symmetry might be the conservation of a single component of the spin operator which can happen in a constant magnetic field if the spin-orbit coupling is weak.

\subsection{Nodal sphere} \label{nss}
Given the set of four-by-four Dirac matrices $\gamma^\mu$ that satisfy the Clifford algebra $\{\gamma^\mu, \gamma^{\nu} \}=2 \eta^{\mu\nu}$ with the Minkowski metric $\eta^{\mu\nu}$ and $\mu, \nu=0,x,y,z$, we first define $\alpha^i=\gamma^0 \gamma^i$ and $\gamma^5=\mathrm{i}\gamma^0 \gamma^x \gamma^y \gamma^z$. Now
consider the Hamiltonian, which describes a massless Dirac fermion perturbed by a $\gamma^5$ term
\beq \label{mod1}
\mathcal{H}(\mathbf{k})=k_i \alpha^i- \lambda \gamma^5,
\eeq
where $k_i$ is the crystal momentum and $\lambda \in \mathbb{R}$. Since $[\gamma^5, \mathcal{H}]=0$, the model has an internal $U(1)$ symmetry \footnote{In high energy physics this is known as the axial symmetry.} generated by the matrix $\gamma^5$.  In this paper we use the chiral representation of the Dirac matrices resulting in $\alpha^i=\sigma^{z}\otimes\tau^{i}$ and $\gamma^{5}=-\sigma^{z} \otimes \tau^{0}$. The last term in  \cref{mod1} preserves time-reversal, but breaks the inversion symmetry. It splits the Dirac point at $\mathbf{k}=0$ into a pair of Weyl points of opposite chirality by separating them in energy by $2\lambda$. This model is in DIII symmetry class of the ten-fold way classification \cite{Schnyder2008}.  It appeared in the context of studies of the chiral magnetic effect \cite{Fukushima2008}.
The energy spectrum
\beq
E(\mathbf{k})=\pm |\mathbf{k}|\pm \lambda
\eeq
exhibits a band degeneracy at the Fermi level $E=0$ on a sphere defined by $|\mathbf{k}|=|\lambda|$.  The nodal surface is protected by the $\gamma^{5}$ symmetry against perturbations since two bands that cross each other have different $\gamma^{5}$ eigenvalues and cannot be hybridized. Under a generic $\gamma^5$-symmetric perturbation the sphere will deform and move away from $E=0$, but will not be gapped out. On the other hand, by adding a mass term $\sim\gamma^0= \sigma^x\otimes \tau^0$ to the Hamiltonian, given by \cref{mod1}, the symmetry is broken and the nodal sphere disappears. This behavior can be understood using topology, as we will explain in the following.

On a zero-dimensional ($d_m=0$) enclosing manifold (Fig. \ref{fig1} a) an integer-valued topological invariant tied to the $\gamma^5$ symmetry can be defined  as  the $\gamma^5$ quantum number of the lower band that generates the nodal sphere (in our model this is the  occupied band with the highest energy). Given a normalized Bloch state $|u(\mathbf{k)}\rangle$ of this band, we define
\beq \label{cof}
c_0(\mathbf{k})=\langle u(\mathbf{k})| \gamma^5 | u(\mathbf{k}) \rangle.
\eeq
If this band is degenerate in energy, the $\gamma^5$ quantum numbers of individual bands should be summed, thus $c_{0}\in\mathbb{Z}$. 
 In the presence of a nodal surface one can define a topological invariant $\Delta c_0=\frac{1}{2}[c_{0}(\mathbf{k}_{\text{in}})-c_{0}(\mathbf{k}_{\text{out}})]$, where the momentum $\mathbf{k}_{\text{in(out)}}$ is  located anywhere inside (outside) the nodal surface. Since the occupied band with the highest energy has opposite $\gamma^5$ quantum numbers outside and inside the nodal surface, the difference $\Delta c_0$ is non-trivial for the model given by \cref{mod1}. As discussed in Sec. \ref{intro},  if we think of a general $\gamma^5$-symmetric system with a nodal sphere, we cannot gap out a nodal sphere with an infinitesimal $\gamma^5$-symmetry preserving term, if  $\Delta c_0\in \mathbb{Z} \setminus  \{ 0\}$. Consequently we conclude that $\Delta c_0$ is a $\mathbb{Z}$-valued topological charge.

In addition, an integer-valued topological invariant can be defined on a two-dimensional ($d_m=2$) manifold enclosing the nodal sphere (Fig. \ref{fig1} c). This is the Chern number of the lower band that generates the nodal sphere (the occupied band with the highest energy)\beq \label{Chc2}
c_2(S)= \frac{\mathrm{i}}{2\pi} \oiint_{S} \D \mathbf{k} \, \nabla_{\mathbf{k}} \times \mathcal{A}(\mathbf{k}), 
\eeq
where the Berry connection $\mathcal{A}(\mathbf{k})=\langle u (\mathbf{k})| \nabla_{\mathbf{k}}u (\mathbf{k})\rangle$. In the model, given by \cref{mod1}, this Chern number is non-trivial but does not change as the radius of the enclosing manifolds crosses the nodal sphere, i.e. $c^{\text{in}}_2=c^{\text{out}}_2$. Since the difference of the Chern numbers \cref{Chc2} inside and outside the nodal sphere is zero, we conclude that the nodal surface in this model has no robustness with respect to a generic perturbation that breaks $\gamma^5$ symmetry. Such perturbation, as for example the mass term $\sim\gamma^0= \sigma^x\otimes \tau^0$, fully gaps out the nodal surface.  

We now investigate the surface states of the model given by \cref{mod1}.
Imagine that the system has a boundary  at $z=0$ and fills only a half of the space $z>0$. Due to $\gamma^5$ symmetry the Hamiltonian splits into two decoupled Weyl blocks. Consequently, one expects two degenerate Fermi arcs starting at the nodal surface that are protected by $\gamma^5$ symmetry. This can be demonstrated analytically by  first introducing the following boundary condition
$
\sigma^0 \otimes \tau^y \psi=\psi
$
that preserves helicity of excitations. In the spirit of \cite{Witten2016}, we extend the boundary condition also into the bulk. Solving the Schr\"odinger equation  $\mathcal{H}\psi_{\mp}=E_{\mp}\psi_{\mp}$, where $\mathcal{H}$ is given by  \cref{mod1} gives the surface states $ \psi_{-}=(0,\phi_{-} )^T$ and  $\psi_{+}=( \phi_+,0)^T$ with
\beq
\begin{split}
&\phi_{\mp}=\exp(\mathrm{i}k_{x}x+\mathrm{i}k_{y}y-\kappa z)\phi^0,
\end{split}
\eeq
and
\begin{equation}
	E_{\mp}=\mp(k_{y}+\lambda),
\end{equation}
where $\phi^0$ is a constant spinor that satisfies $ \tau^y \phi^0=\phi^0$ and $\kappa =-k_{x}$.
For $\kappa>0$ these solutions are normalizable, and localized close to the boundary. If we set $E=0$, we find a doubly-degenerate Fermi arc at the surface BZ  at $(k_x<0, k_y=-\lambda)$. It starts at ($k_{x}=0$), i.e. the boundary of the projection of the nodal surface on the surface BZ (see \cref{fig: fig2}).  $\psi_{+}$ and $\psi_{-}$   have opposite $\gamma^5$ eigenvalues and counter-propagate along the $y$ axis. 
These counter-propagating surface modes are robust under $\gamma^5$-symmetric perturbations $\mathcal{H}_{\text{imp}}$ since $\langle\psi_{-}|\mathcal{H}_{\text{imp}}|\psi_{+}\rangle=0$.

After calculating the  surface states, we investigate their topologically stability. We have seen that each sub-block of  \cref{mod1} produces its
own surface state with a particular $\gamma_{5}$ quantum number.
As long as $\gamma_{5}$ quantum number is conserved,  states with different quantum numbers cannot influence each other.
Thus we have to check the topological properties of each subsystem
with a definite $\gamma_{5}$ eigenvalue, separately. However, those subsystems are just Weyl Hamiltonians \emph{displaced}
in energy. Therefore, what we are actually interested in is, the topology
of \emph{Fermi surfaces} of the  Weyl sub-systems. The appropriate topological invariant for the blocks is  $\gamma^5$-Chern number, which is given by
\begin{equation}
c_2^{(\ell)}(S)=\sum_{n\in \operatorname{occupied}}\frac{\mathrm{i}}{2\pi} \oiint _{S} \D \mathbf{k}\langle\nabla_{\mathbf{k}}u_{n,\ell}(\mathbf{k})|\times|\nabla_{\mathbf{k}}u_{n,\ell}(\mathbf{k})\rangle\,\label{eq: g5cn}
\end{equation}
where $\ell$ denotes the eigenvalue of $\gamma^{5}$ matrix.
 We emphasize that $c_2^{(\ell)}$ characterises the stability of the surfaces states under $\gamma^{5}$ preserving infinitesimal perturbation, on the other hand the Chern number of the  upper filled band \cref{Chc2} characterizes the stability of the nodal surface under generic infinitesimal perturbations. For our model, the absolute value of the $\gamma^{5}$-Chern numbers is unity (zero) for enclosing manifolds of radius larger (smaller) than the radius of the nodal sphere.

\begin{figure}[ht]
\begin{center}
\includegraphics[width=0.48\textwidth]{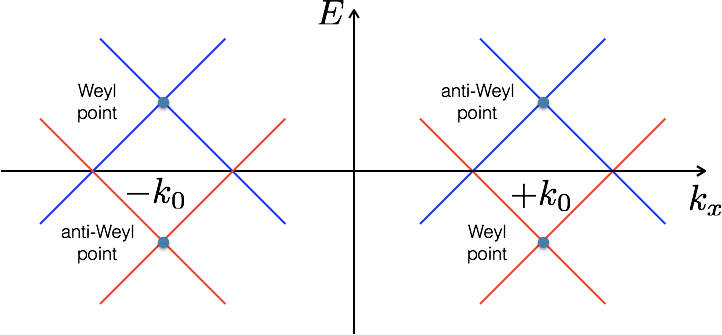}
\caption{Schematic spectrum of the lattice model \cref{mod1latt} for $t, \, \lambda>0$: Red and blue bands have $\gamma^5$ quantum number $+1$ and $-1$ and cannot be hybridized and connect to each other.}\label{figdoub2}
\end{center}
\end{figure}

\begin{figure}[ht]
\begin{center}
\includegraphics[width=0.48\textwidth]{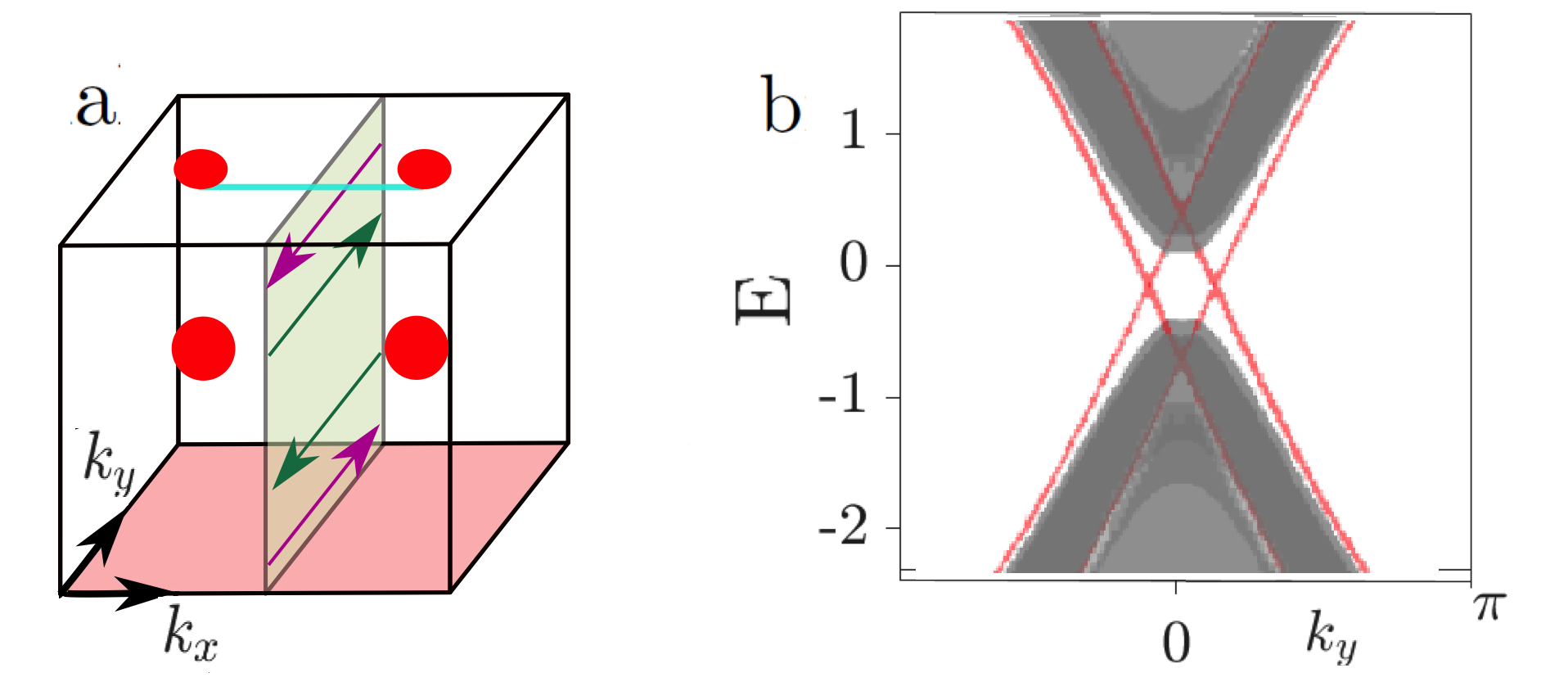}
\caption{Fermi arcs in a slab geometry: (a)A degenerate pair of counter-propagating surface Fermi arcs (cyan) on the upper boundary connecting the projections of nodal Fermi surfaces. For simplicity the Fermi arcs on the lower boundary are not displayed; (b) Fermi arcs (red) in the energy spectrum of the Hamiltonian \cref{mod1latt}. The counter-propagating  surface states has opposite $\gamma^5$ quantum numbers and cannot be hybridized by symmetry-preserving perturbations.  }\label{fig: fig2}
\end{center}
\end{figure}

The continuum model \cref{mod1} cannot be used in the full BZ since it violates its periodicity. We conclude this section with a discussion of a minimal lattice realization of this model. The  lattice Hamiltonian is given by
\beq \label{mod1latt}
\begin{split}
\mathcal{H}(\mathbf{k})=&-\Big([2-\cos k_y-\cos k_z]+2 t [\cos k_x-\cos k_0] \Big)\alpha^x \\
&-2 t \sin k_y \alpha^y -2 t \sin k_z \alpha^z- \lambda \gamma^5,
\end{split}
\eeq
where $t\in\mathbb{R}$.
The model originates from the minimal lattice model of Weyl semimetals introduced in  \cite{Yang2011} and discussed extensively in \cite{McCormick2017}. As sketched in Fig. \ref{figdoub2}, the lattice model has two pairs of Weyl points located at $\mathbf{k}=(\pm k_0, 0, 0)$. Due to the $U(1)$ symmetry only Weyl points of the same $\gamma^5$  quantum number can be connected to each other. Hence the Nielsen-Ninomiya theorem must be applied in the two $\gamma^5$ sectors separately and nodal spheres appear necessarily in pairs.

Since there are two nodal surfaces in the minimal lattice realization, it is natural to expect that the Fermi arcs that we found above will connect the two nodal surfaces projected on the boundary BZ (see Fig. \ref{fig: fig2} a). This is indeed what one finds by diagonalizing numerically the lattice Hamiltonian in a slab geometry (see Fig. \ref{fig: fig2} b). The Fermi arcs are protected by the $\gamma^5$ symmetry since by adding a mass term $\sim \sigma^x\otimes \tau^0$ to the lattice Hamiltonian \cref{mod1latt}, the Fermi arcs hybridize and gap out.

\subsection{Nodal torus} \label{ntu}
We start from a four-band model with the effective low-energy Hamiltonian
\beq \label{mod2}
\mathcal{H}(\mathbf{k})=(k_{\perp}-k_0) \alpha^x+ k_z \alpha^y - \lambda \gamma^5,
\eeq
with $k_{\perp}=\sqrt{k_{x} ^{2}+k_{y} ^{2}}$. This model has the $\gamma^5$ symmetry. The last term in  \cref{mod2} splits a doubly-degenerate (Dirac) loop of radius $k_0$ at $k_z=0$ into a pair of Weyl loops by separating them in energy by an amount of $2\lambda$.
The dispersion relation for this system is given by
\beq
E(\mathbf{k})=\pm \sqrt{(k_{\perp}-k_{0})^{2}+k_{z} ^{2}}\pm\lambda,
\eeq
and exhibits a band degeneracy at $E=0$ with $\sqrt{(k_{\perp}-k_{0})^{2}+k_{z} ^{2}}=\lambda$ which for $\lambda<k_0$ forms a torus in momentum space.

In a close analogy to the nodal sphere discussed in Sec. \ref{nss},  the nodal torus is protected by the unitary $\gamma^5$ symmetry. The appropriate $\mathbb{Z}$-valued topological invariant defined on a zero-dimensional enclosing manifold is given by  \cref{cof}. For the model \cref{mod2} the difference of the invariants defined inside and outside the nodal torus is non-trivial which ensures its full robustness under  infinitesimal  $\gamma^5$-preserving perturbations. 

In addition to the $\gamma^5$ symmetry, the model, given by  \cref{mod2}, is also invariant under the combination of inversion $P$ and  time reversal $T$, i.e. $PT$\footnote{\label{fn1}. In addition to the $PT$ symmetry, the nodal tori models \cref{mod2} and \cref{mod4} are also invariant a unitary mirror ($z\to -z$) symmetry, but we do not discuss its consequences in this paper.},  which implies the following constraint on the   Hamiltonian 
\beq
U_{PT} \mathcal{H}^{*}(\mathbf{k})U_{PT}^{-1}=\mathcal{H}(\mathbf{k}),
\eeq
where $U_{PT}=\alpha^x$.
Analogously to the previous section, we can treat the diagonal sub-blocks of this system separately. The sub-blocks give rise nodal loops  lying at energies $\pm\lambda$, 
which are protected by the $PT$ symmetry. The topological invariant that characterizes the stability  of such nodal loops is the Berry phase which is quantized in  $PT$-symmetric systems\footnote{If we break $PT$ symmetry, the block Berry phase can be  changed continuously.}\cite{rui2017}. In our model \cref{mod2} we find $\pi$ Berry phase for the blocks.
In addition, the Berry curvature is forced to be  zero for every non-singular point in the BZ. As a result, $\pi$ Berry phase guarantees the existence of nodal loops that
 are stable   under infinitesimal perturbations that preserve  $PT$ and $\gamma^5$ symmetries. On the other hand, the $PT$ symmetry is not required for the stability of the nodal torus.  In fact, if we break this symmetry without breaking $\gamma^5$ symmetry, the nodal torus will not be destroyed, but the nodal loops lying at   energies $\pm\lambda$ will be gapped. Hence we can refer to $PT$ symmetry as an accidental symmetry. 

\begin{figure}[h]
	\includegraphics[width=0.47\textwidth]{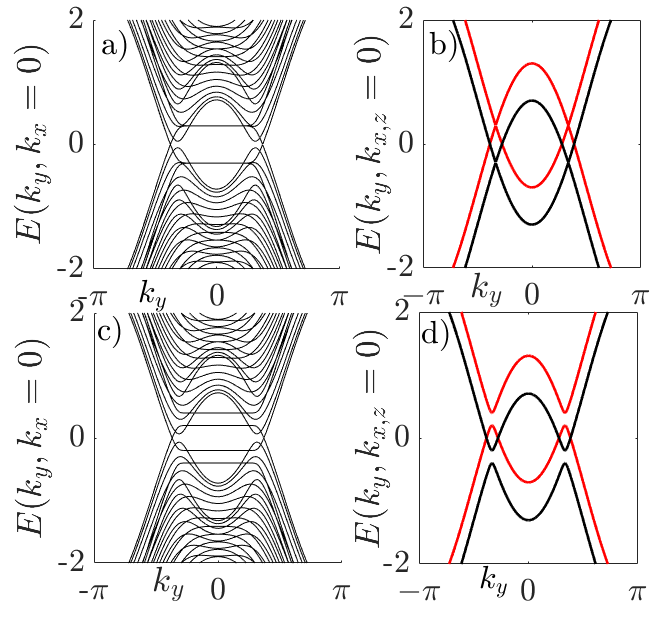}
	\caption{\label{fig: nodalloops} Energy spectrum of the model \eqref{mod2} in the (a) slab and (b) infinite geometry. The drumhead surface states in (a)  lie at the energy levels of the nodal loops. In (b) the crossings of the red and black bands  at the Fermi level form a nodal torus. The panels (c) and (d) display the energy spectrum of the model \eqref{mod2} perturbed by a term that breaks accidental $PT$ symmetry, but preserves $\gamma^5$ symmetry.}
\end{figure}

In order to illustrate the arguments given above in a concrete example, we consider a minimal lattice model that hosts a nodal torus. Its Hamiltonian is given by 
\beq \label{mo21latt} 
\begin{split}
\mathcal{H}(\mathbf{k})=&-\Big(6t_1-2 t_2 [\cos k_x+ \cos k_y+ \cos k_z] \Big)\alpha^x \\
&-2 t_2 \sin k_z \alpha^y- \lambda \gamma^5,
\end{split}
\eeq
where $t_{1,2}$ are constants which fix the radius of the torus. This Hamiltonian  is motivated by the lattice model of a nodal loop semimetal investigated in \cite{Wang2017}. It is worth pointing out that this lattice model has just  one nodal torus in the BZ  (no Nielsen-Ninomiya doubling).
The bulk energy spectrum of this Hamiltonian is shown in  \cref{fig: nodalloops}-b. 
Furthermore, in a slab geometry (as a result of the existence of the nodal loops in  the bulk energy spectrum) we find  drumhead surface states that have the same energy as the nodal loops as shown in \cref{fig: nodalloops}-a. A $\gamma^5$-invariant perturbation that breaks the $PT$ symmetry destroys the nodal loops, but not the nodal torus (see \cref{fig: nodalloops}-c,d).

\section{Nodal surfaces protected by anti-unitary symmetry} \label{nsaU}
Here we turn to four-band models that exhibit nodal surfaces which are protected by anti-unitary symmetries. These objects appeared in the recent literature \cite{Agterberg2017,  Bzdusek2017, Timm2017} and formed the starting point of our investigation. 

\subsection{Nodal sphere} \label{nst}
Consider a four-band model with the Hamiltonian
\beq \label{infW}
\mathcal{H}(\mathbf{k})= k_i \tilde \alpha^i- \lambda \gamma^5-\delta \gamma^0,
\eeq
where  $\tilde\alpha^i=\sigma^0\otimes \tau^i$, $\gamma^5=-\sigma^z\otimes \tau^0$ and $\gamma^0=\sigma^x\otimes \tau^0$. 
The model consists of a pair of Weyl points of the same chirality which are split in energy by the last two terms in  \cref{infW}. Notice that the Hamiltonian has no unitary $U(1)$ symmetry apart from the particle number conservation.
Even in the absence of a unitary symmetry, the spectrum of the Hamiltonian \cref{infW}
\beq
E(\mathbf{k})=\pm |\mathbf{k}|\pm \sqrt{\lambda^2+\delta^2}
\eeq
contains a band degeneracy at $E=0$ which  is located on a sphere of radius $|\mathbf{k}|=\sqrt{\lambda^2+\delta^2}$.

We will argue now that in this model the nodal surface is protected by an anti-unitary combination of inversion $P$ and particle-hole $C$ symmetry \cite{Kobayashi2014, Zhao2016} and construct a $\mathbb{Z}_2$-valued topological invariant tied to this symmetry \cite{Agterberg2017, Bzdusek2017}.
To this end consider a class of Hamiltonians of the form
\beq \label{PCH}
\mathcal{H}(\mathbf{k})=a_i (\mathbf{k}) \tilde \alpha^i+ b_i (\mathbf{k})\tilde\beta^i
\eeq
with $\tilde\alpha^i=\sigma^0\otimes \tau^i$ and $\tilde\beta^i=\sigma^i\otimes \tau^0$ and $\mathbf{a}$ and $\mathbf{b}$ being real functions of $\mathbf{k}$. The model \cref{infW} is a special case of \cref{PCH} since $\gamma^0 =\tilde \beta^x$ and $\gamma^5=-\tilde \beta^z$. It will turn out important in the following that the Hamiltonian \cref{PCH} can be unitary rotated to an antisymmetric form \cite{Agterberg2017, Bzdusek2017} to be denoted by $\bar{\mathcal{H}}$.   The spectrum of \cref{PCH}
\beq
E(\mathbf{k})=\pm|\mathbf{a}(\mathbf{k})|\pm|\mathbf{b}(\mathbf{k})|
\eeq
supports a zero-energy nodal surface at $|\mathbf{a}(\mathbf{k})|=|\mathbf{b}(\mathbf{k})|$.
A generic model \cref{PCH} breaks inversion $P$, time-reversal $T$ and particle-hole $C$ symmetries. Nevertheless, it is invariant under the anti-unitary $PC$ symmetry which acts on the Hamiltonian as
\beq
U_{PC} \mathcal{H}^{*}(\mathbf{k})U_{PC}^{-1}=-\mathcal{H}(\mathbf{k}).
\eeq
In our representation $U_{PC}=\sigma^y\otimes \tau^y$ is real. Note that $(PC)^2=U_{PC}U_{PC}^*=+\mathbb{1}$. Due to this symmetry the nodal surface is fixed to the Fermi level and thus necessarily coincides with the Fermi surfaces of the two touching bands. It was found in \cite{Kobayashi2014, Zhao2016} that three-dimensional Fermi surfaces with $PC$ symmetry that squares to unity  have topological $\mathbb{Z}_2$ classification. In our formulation, the topological invariant defined on a zero-dimensional enclosing manifold can be calculated as the sign of the Pfaffian of the antisymmetric form of the Hamiltonian $\bar{\mathcal{H}}$
\beq \label{pfc}
c_0(\mathbf{k})=\text{sgn}\text{Pf} \, \bar{\mathcal{H}}(\mathbf{k}).
\eeq

Since in our representation the Hamiltonian \cref{PCH} is not antisymmetric, in order to compute the invariant, we must first unitary rotate it to the antisymmetric form 
$
\bar{ \mathcal{H}}= \Omega \mathcal{H} \Omega^\dagger.
$
The resulting Pfaffian 
\beq \label{Pfeq}
\text{Pf} \, \bar {\mathcal{H}}=\mathbf{b}^2(\mathbf{k})-\mathbf{a}^2(\mathbf{k})
\eeq
is real and vanishes on the nodal surface.
In particular, for the model \cref{infW} the Pfaffian has a simple zero on the nodal sphere and thus it has opposite signs inside and outside of it. Following the arguments of Sec. \ref{intro}, the non-trivial difference of the Pfaffians makes the nodal sphere fully robust with respect to small $PC$-invariant perturbations. We notice that although the choice of $\Omega$ and the sign of the resulting $c_0(\mathbf{k})$ is not unique, this ambiguity does not affect the difference of the topological invariants.

We discuss now the Chern number invariant \cref{Chc2} in the context of the model \cref{infW}.
 For an enclosing manifold located inside (outside) the nodal sphere the band Chern number is non-trivial and in addition $c^{\text{in}}_2=-c^{\text{out}}_2$. The non-trivial difference of the band Chern numbers $\Delta c_2=c^{\text{in}}_2-c^{\text{out}}_2$ ensures that a generic small perturbation cannot fully gap out the nodal sphere, but has to leave in the band structure at least a pair of Weyl points close to the Fermi level. Thus contrary to the nodal objects from Sec. \ref{nodalU1}, the nodal sphere discussed here has certain robustness with respect to arbitrary perturbations of the Hamiltonian.
 
In the presence of a spatial boundary, for $\lambda=\delta=0$ in  \cref{infW} a pair of same-chirality Weyl points at the Fermi level gives rise to a pair of chiral co-propagating zero-energy Fermi arcs. As the Weyl points are split in energy  for $\lambda, \, \delta \ne 0$, the Fermi arcs survive but must start at the Fermi surfaces. These surface states are robust against arbitrary small perturbations thanks to the the total Chern number
\beq \label{Chern}
\text{total Chern number}=\frac{i}{2\pi} \int_{S^2} d \mathbf{k} \, \text{tr} \mathcal{F}
\eeq
which is non-trivial on an enclosing manifold outside the Fermi surfaces. Here the Berry curvature two-form $\mathcal{F}=d \mathcal{A}+\mathcal{A}\wedge \mathcal{A}$ is defined in terms of the non-abelian Berry connection $\mathcal{A}^{ab}=\langle u^a(\mathbf{k})| d u^b(\mathbf{k}) \rangle$, where $a,b$ label only occupied bands. On an enclosing manifold living inside the nodal sphere the total Chern number \cref{Chern} vanishes and thus the nodal sphere represents a locus of the total Berry flux. It is an inflated double Weyl monopole \cite{Agterberg2017}.

Due to the Nielsen-Ninomiya theorem the nodal spheres protected by the anti-unitary $PC$ symmetry should always appear in pairs. To demonstrate it explicitly we considered the minimal lattice Hamiltonian
\beq \label{latdoub}
\begin{split}
\mathcal{H}(\mathbf{k})=&-\Big([2-\cos k_y-\cos k_z]+2 t [\cos k_x-\cos k_0] \Big)\tilde \alpha^x \\
&-2 t \sin k_y \tilde \alpha^y -2 t \sin k_z \tilde \alpha^z-\delta \gamma^0- \lambda \gamma^5.
\end{split}
\eeq
and determined numerically its energy spectrum in the bulk.
In addition, we examined the energy spectrum of this lattice Hamiltonian in a slab geometry. As expected, it contains a pair of chiral zero-energy Fermi arcs which are robust with respect to arbitrary perturbations of the Hamiltonian.

\subsection{Nodal torus} \label{ntt}
Finally we construct and analyze a four-band model that exhibits a nodal torus that is protected by an anti-unitary symmetry.  The model is defined by the real Hamiltonian
\beq \label{mod4}
\mathcal{H}(\mathbf{k})=(k_{\perp}-k_0) \tilde \alpha^x+ k_z \tilde \alpha^z - \lambda \gamma^5-\delta \gamma^0
\eeq
with $k_{\perp}=\sqrt{k_{x} ^{2}+k_{y} ^{2}}$. Apart from the particle number symmetry, there is no $U(1)$ symmetry. Nevertheless the energy spectrum
\beq
E(\mathbf{k})=\pm \sqrt{(k_{\perp}-k_{0})^{2}+k_{z} ^{2}}\pm\sqrt{\lambda^2+\delta^2}
\eeq
has a zero-energy band degeneracy at $\sqrt{(k_{\perp}-k_{0})^{2}+k_{z} ^{2}}=\sqrt{\lambda^2+\delta^2}$ forming a torus in momentum space for $\sqrt{\lambda^2+\delta^2}<k_0$. One can view  the nodal torus as an inflated double Weyl loop \cite{Agterberg2017}.

Since the model \cref{mod4} falls into the class of $PC$-symmetric Hamiltonians \cref{PCH}, similar to the nodal sphere discussed in section \ref{nst}, one can define the Pfaffian $\mathbb{Z}_2$ invariant \cref{pfc}. By evaluating the Pfaffian \cref{Pfeq} in the model \cref{mod4}, one finds that the difference of the Pfaffian invariant inside and outside the nodal torus is non-trivial and thus this object cannot be gapped out by small $PC$-invariant perturbations of the Hamiltonian.


Incidentally, the model \cref{mod4} enjoys more symmetries. Similar to the nodal torus model discussed in \cref{ntu}, this system has the anti-unitary $PT$ symmetry. In this case the Hamiltonian is real and $PT$ acts on the Hamiltonian as
\beq \label{PTsym}
\mathcal{H}^{*}({\mathbf{k}})=\mathcal{H}({\mathbf{k}}).
\eeq
In general, this symmetry implies that the Chern number \cref{Chern} is zero in this symmetry class.
In addition, this symmetry together with the anti-unitary $PC$ symmetry implies the unitary chiral sublattice (anti)symmetry of the model \cref{mod4}
\beq
\{\mathcal{H}(\mathbf{k}), U_S \}=0, \qquad U_S=\sigma^y \otimes \tau^y
\eeq
and gives rise to a $\mathbb{Z}_2$ winding number topological invariant introduced in \cite{Bzdusek2017}. In order to construct this invariant, it is convenient to perform a unitary rotation $\Omega$ which diagonalizes the chiral sublattice symmetry operator $S\to\Omega^\dagger S \Omega=\sigma^z\otimes \tau^0$. This transformation brings the Hamiltonian into the block off-diagonal form
\beq \label{chH}
\mathcal{H}(\mathbf{k})=\left(
\begin{array}{cc}
 0 & h(\mathbf{k}) \\
 {h}^\dagger(\mathbf{k}) & 0
\end{array}
\right),
\eeq
Here due to the $PT$ symmetry the block Hamiltonian $h$ is real. 
To proceed, it is useful to define a flattened Hamiltonian which for a general non-interacting fermionic system with $n$ filled and $m$ empty bands reads
\beq
Q(\mathbf{k})=U(\mathbf{k})\left(
\begin{array}{cc}
 \mathbb{1}_{m\times m} & 0 \\
0 & -\mathbb{1}_{n\times n}
\end{array}
\right)
U^{\dagger}(\mathbf{k}),
\eeq
where the unitary matrix $U(\mathbf{k})$ diagonalizes the Hamiltonian
\beq
\begin{split}
&U^{\dagger}(\mathbf{k}) \mathcal{H}(\mathbf{k}) U(\mathbf{k})= \\
&=\text{diag}(\underbrace{\epsilon_{m+n}(\mathbf{k}), \dots, \epsilon_{n+1}(\mathbf{k})}_{\text{empty bands}}, \underbrace{\epsilon_{n}(\mathbf{k}), \dots, \epsilon_{1}(\mathbf{k})}_{\text{filled bands}}).
\end{split}
\eeq
The flattened Hamiltonian $Q(\mathbf{k})$ is well-defined only away from band degeneracies. In the presence of the chiral sublattice symmetry $m=n$ and the flattened Hamiltonian is block off-diagonal
\beq \label{chq}
Q(\mathbf{k})=\left(
\begin{array}{cc}
 0 & q(\mathbf{k}) \\
q^\dagger(\mathbf{k}) & 0
\end{array}
\right)
\eeq
in any basis which diagonalizes the operator $S$.
Note that in general $q(\mathbf{k})\in \text{U}(n)$ since $Q^2=\mathbb{1}$. The additional reality condition \cref{PTsym} that follows from the $PT$ symmetry implies that the flattened block $q(\mathbf{k})\in \text{O}(n)$. 
In particular, for the four-band model \cref{mod4} one has $q(\mathbf{k})\in \text{O}(2)$.
Now  on any closed one-dimensional manifold that does not intersect the torus in the BZ one can compute the topological winding number \cite{Bzdusek2017} as the homotopy equivalence class of mappings $S^1\to SO(2)$. We define the winding number as
\beq
c_1=-\frac{i}{4\pi}\oint ds \text{Tr} \, \big[ q^{T}\sigma^y \partial_s q \big],
\eeq
where $s$ is a coordinate that parametrizes the one-dimensional enclosing manifold. The invariant
 is integer-valued since the homotopy group $\pi_1[SO(2)]=\mathbb{Z}$ \footnote{Since $\pi_1[SO(n)]=\mathbb{Z}_2$ for $n>2$, the winding number is only $\mathbb{Z}_2$-valued for models with more than four bands \cite{Bzdusek2017}.}. It is instructive now to evaluate the winding number invariant for the nodal torus model \cref{mod4}. To this end we consider an enclosing $S^1$ manifold of radius $k_S$ that is positioned in the $x-z$ plane and is centered at the momentum $\mathbf{k}=(k_0, 0, 0)$. A straightforward calculation reveals that if $k_S>\sqrt{\lambda^2+\delta^2}$, i.e., the manifold encloses the torus from outside and links with it, the absolute value of the winding number $c^{\text{out}}_1$ is equal to unity. On the other hand for $k_S<\sqrt{\lambda^2+\delta^2}$, the winding number $c^{\text{in}}_1$ vanishes since in this case the enclosing manifold can be shrunk to a point without crossing energy bands. Following general arguments from Sec. \ref{intro}, the non-trivial difference of the winding numbers ensures that the nodal torus cannot be fully gapped out by small perturbations that are invariant under $PC$ and $PT$ symmetries. Note however that in the present case this prediction has little practical value because the nodal torus is fully protected against any small $PC$-symmetric perturbation by the Pfaffian invariant. 
It would be interesting to find a model protected by the non-trivial difference of a topological invariant defined on one-dimensional enclosing manifolds, where a nodal surface is gapped out to nodal loop(s).

We observe that similar to the model discussed in Sec. \ref{ntu}, the present model does not support robust zero-energy surface states. The pair of zero-energy drumhead surface states present in the limit of the double-Weyl loop ($\lambda=\delta=0$) is shifted to a finite energy by the last two terms in the Hamiltonian \cref{mod4}. One arrives to the same conclusion by diagonalizing the lattice the Hamiltonian 
\beq \label{mo21latta}
\begin{split}
\mathcal{H}(\mathbf{k})=&-\Big(6t_1-2 t_2 \left[ \cos k_x+ \cos k_y+ \cos k_z \right] \Big)\tilde\alpha^x \\
&-2 t_2 \sin k_z \tilde\alpha^z- \lambda \gamma^5-\delta \gamma^0
\end{split}
\eeq
in a slab geometry.

%

\section{Conclusion and outlook}
In this paper we investigated the physics of three-dimensional fermionic band models that exhibit two-dimensional Weyl nodal surfaces. We argued that the robustness of these nodal objects is ensured by topology. Specifically, we showed that the degree of robustness is determined by the dimensionality of gapped enclosing manifolds, where topological invariants are evaluated.
We demonstrated this in several four-band toy models that exhibit nodal surfaces protected by unitary or anti-unitary symmetry. It would be interesting to study the effects of interactions and disorder on nodal surfaces, investigate transport properties and find realistic models of materials where these ideas can be applied.

\section*{Acknowledgments:}
We acknowledge fruitful discussions with Barry Bradlyn, Anton Burkov, Tom\'a\v s Bzdu\v sek, Titus Neupert and Grigori Volovik. Our work is supported by the Emmy Noether Programme of  German Research Foundation (DFG) under grant No. MO 3013/1-1.

\appendix
\section{Topological robustness of general nodal surfaces}\label{App1}
\begin{figure}[ht]
\begin{center}
\includegraphics[height=0.35\textwidth]{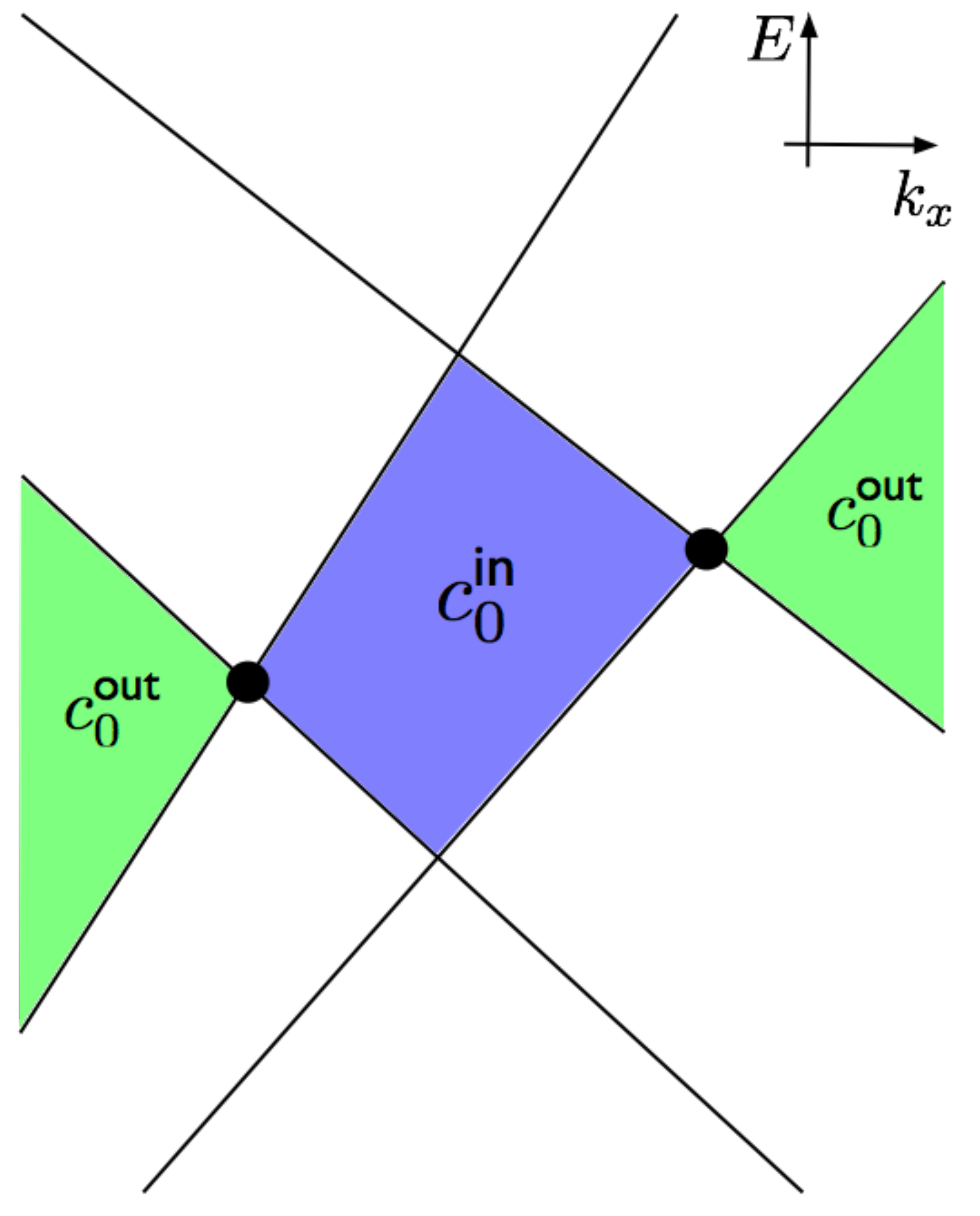}
\caption{A  two-dimensional cut of the four-dimensional energy-momentum space partitioned by bands (black solid lines) into regions. The nodal surface projects as two black dots on the cut. Nontrivial difference of topological invariants $c_0^{\text{out}}$ and $c_0^{\text{in}}$ defined on any two zero-dimensional point manifolds located within the outer (green) and inner (blue) regions, respectively, protects the nodal surface from being gapped. }\label{fig1a}
\end{center}
\end{figure}

Although in the main part of the paper we limit our attention to nodal surfaces tuned to the Fermi level, this assumption is not necessary and general (dispersing in energy) nodal surfaces exhibit topological robustness to small perturbations in the way we put forward in Sec. \ref{intro}.
To understand the dispersing case, it is useful to consider a four-dimensional energy-momentum space which is naturally partitioned by three-dimensional energy bands into four-dimensional regions. If a topological invariant of the band structure can be defined on a manifold residing in some four-dimensional region, by construction it must be constant within the given region. In general, a non-trivial difference of topological invariants  evaluated  on enclosing manifolds of dimensions $d_{\text{m}}=0,1,2$ embedded in different properly chosen regions of the energy-momentum space guarantees a nodal object of dimension $d=2,1,0$ in the presence of any small perturbation compatible with the given topological invariant. For the case $d_{\text{m}}=0$, the mechanism is illustrated in Fig. \ref{fig1a}.

\newpage

\bibliography{library}

\begin{thebibliography}{38}%
\makeatletter
\providecommand \@ifxundefined [1]{%
 \@ifx{#1\undefined}
}%
\providecommand \@ifnum [1]{%
 \ifnum #1\expandafter \@firstoftwo
 \else \expandafter \@secondoftwo
 \fi
}%
\providecommand \@ifx [1]{%
 \ifx #1\expandafter \@firstoftwo
 \else \expandafter \@secondoftwo
 \fi
}%
\providecommand \natexlab [1]{#1}%
\providecommand \enquote  [1]{``#1''}%
\providecommand \bibnamefont  [1]{#1}%
\providecommand \bibfnamefont [1]{#1}%
\providecommand \citenamefont [1]{#1}%
\providecommand \href@noop [0]{\@secondoftwo}%
\providecommand \href [0]{\begingroup \@sanitize@url \@href}%
\providecommand \@href[1]{\@@startlink{#1}\@@href}%
\providecommand \@@href[1]{\endgroup#1\@@endlink}%
\providecommand \@sanitize@url [0]{\catcode `\\12\catcode `\$12\catcode
  `\&12\catcode `\#12\catcode `\^12\catcode `\_12\catcode `\%12\relax}%
\providecommand \@@startlink[1]{}%
\providecommand \@@endlink[0]{}%
\providecommand \url  [0]{\begingroup\@sanitize@url \@url }%
\providecommand \@url [1]{\endgroup\@href {#1}{\urlprefix }}%
\providecommand \urlprefix  [0]{URL }%
\providecommand \Eprint [0]{\href }%
\providecommand \doibase [0]{http://dx.doi.org/}%
\providecommand \selectlanguage [0]{\@gobble}%
\providecommand \bibinfo  [0]{\@secondoftwo}%
\providecommand \bibfield  [0]{\@secondoftwo}%
\providecommand \translation [1]{[#1]}%
\providecommand \BibitemOpen [0]{}%
\providecommand \bibitemStop [0]{}%
\providecommand \bibitemNoStop [0]{.\EOS\space}%
\providecommand \EOS [0]{\spacefactor3000\relax}%
\providecommand \BibitemShut  [1]{\csname bibitem#1\endcsname}%
\let\auto@bib@innerbib\@empty
\bibitem [{\citenamefont {Hasan}\ and\ \citenamefont {Kane}(2010)}]{Hasan2010}%
  \BibitemOpen
  \bibfield  {author} {\bibinfo {author} {\bibfnamefont {M.~Z.}\ \bibnamefont
  {Hasan}}\ and\ \bibinfo {author} {\bibfnamefont {C.~L.}\ \bibnamefont
  {Kane}},\ }\href {\doibase 10.1103/RevModPhys.82.3045} {\bibfield  {journal}
  {\bibinfo  {journal} {Rev. Mod. Phys.}\ }\textbf {\bibinfo {volume} {82}},\
  \bibinfo {pages} {3045} (\bibinfo {year} {2010})}\BibitemShut {NoStop}%
\bibitem [{\citenamefont {Qi}\ and\ \citenamefont {Zhang}(2011)}]{Qi2011}%
  \BibitemOpen
  \bibfield  {author} {\bibinfo {author} {\bibfnamefont {X.-L.}\ \bibnamefont
  {Qi}}\ and\ \bibinfo {author} {\bibfnamefont {S.-C.}\ \bibnamefont {Zhang}},\
  }\href {\doibase 10.1103/RevModPhys.83.1057} {\bibfield  {journal} {\bibinfo
  {journal} {Rev. Mod. Phys.}\ }\textbf {\bibinfo {volume} {83}},\ \bibinfo
  {pages} {1057} (\bibinfo {year} {2011})}\BibitemShut {NoStop}%
\bibitem [{\citenamefont {Schnyder}\ \emph {et~al.}(2008)\citenamefont
  {Schnyder}, \citenamefont {Ryu}, \citenamefont {Furusaki},\ and\
  \citenamefont {Ludwig}}]{Schnyder2008}%
  \BibitemOpen
  \bibfield  {author} {\bibinfo {author} {\bibfnamefont {A.~P.}\ \bibnamefont
  {Schnyder}}, \bibinfo {author} {\bibfnamefont {S.}~\bibnamefont {Ryu}},
  \bibinfo {author} {\bibfnamefont {A.}~\bibnamefont {Furusaki}}, \ and\
  \bibinfo {author} {\bibfnamefont {A.~W.~W.}\ \bibnamefont {Ludwig}},\ }\href
  {\doibase 10.1103/PhysRevB.78.195125} {\bibfield  {journal} {\bibinfo
  {journal} {Phys. Rev. B}\ }\textbf {\bibinfo {volume} {78}},\ \bibinfo
  {pages} {195125} (\bibinfo {year} {2008})}\BibitemShut {NoStop}%
\bibitem [{\citenamefont {{Turner}}\ and\ \citenamefont
  {{Vishwanath}}()}]{Turner2013}%
  \BibitemOpen
  \bibfield  {author} {\bibinfo {author} {\bibfnamefont {A.~M.}\ \bibnamefont
  {{Turner}}}\ and\ \bibinfo {author} {\bibfnamefont {A.}~\bibnamefont
  {{Vishwanath}}},\ }\href@noop {} {\ }\Eprint {http://arxiv.org/abs/1301.0330}
  {arXiv:1301.0330} \BibitemShut {NoStop}%
\bibitem [{\citenamefont {Burkov}(2015)}]{Burkov2015}%
  \BibitemOpen
  \bibfield  {author} {\bibinfo {author} {\bibfnamefont {A.}~\bibnamefont
  {Burkov}},\ }\href@noop {} {\bibfield  {journal} {\bibinfo  {journal}
  {Journal of Physics: Condensed Matter}\ }\textbf {\bibinfo {volume} {27}},\
  \bibinfo {pages} {113201} (\bibinfo {year} {2015})}\BibitemShut {NoStop}%
\bibitem [{\citenamefont {{Armitage}}\ \emph {et~al.}()\citenamefont
  {{Armitage}}, \citenamefont {{Mele}},\ and\ \citenamefont
  {{Vishwanath}}}]{Armitage2017}%
  \BibitemOpen
  \bibfield  {author} {\bibinfo {author} {\bibfnamefont {N.~P.}\ \bibnamefont
  {{Armitage}}}, \bibinfo {author} {\bibfnamefont {E.~J.}\ \bibnamefont
  {{Mele}}}, \ and\ \bibinfo {author} {\bibfnamefont {A.}~\bibnamefont
  {{Vishwanath}}},\ }\href@noop {} {\ }\Eprint
  {http://arxiv.org/abs/1705.01111} {arXiv:1705.01111} \BibitemShut {NoStop}%
\bibitem [{\citenamefont {Volovik}(1992)}]{volovik1992exotic}%
  \BibitemOpen
  \bibfield  {author} {\bibinfo {author} {\bibfnamefont {G.~E.}\ \bibnamefont
  {Volovik}},\ }\href@noop {} {\emph {\bibinfo {title} {{Exotic properties of
  superfluid 3He}}}},\ Vol.~\bibinfo {volume} {1}\ (\bibinfo  {publisher}
  {World Scientific},\ \bibinfo {year} {1992})\BibitemShut {NoStop}%
\bibitem [{\citenamefont {Volovik}(2009)}]{volovikbook}%
  \BibitemOpen
  \bibfield  {author} {\bibinfo {author} {\bibfnamefont {G.~E.}\ \bibnamefont
  {Volovik}},\ }\href@noop {} {\emph {\bibinfo {title} {{The universe in a
  helium droplet}}}},\ Vol.\ \bibinfo {volume} {117}\ (\bibinfo  {publisher}
  {Oxford University Press New York},\ \bibinfo {year} {2009})\BibitemShut
  {NoStop}%
\bibitem [{\citenamefont {Xu}\ \emph {et~al.}(2015)\citenamefont {Xu},
  \citenamefont {Belopolski}, \citenamefont {Alidoust}, \citenamefont
  {Neupane}, \citenamefont {Bian}, \citenamefont {Zhang}, \citenamefont
  {Sankar}, \citenamefont {Chang}, \citenamefont {Yuan}, \citenamefont {Lee}
  \emph {et~al.}}]{Xu2015}%
  \BibitemOpen
  \bibfield  {author} {\bibinfo {author} {\bibfnamefont {S.-Y.}\ \bibnamefont
  {Xu}}, \bibinfo {author} {\bibfnamefont {I.}~\bibnamefont {Belopolski}},
  \bibinfo {author} {\bibfnamefont {N.}~\bibnamefont {Alidoust}}, \bibinfo
  {author} {\bibfnamefont {M.}~\bibnamefont {Neupane}}, \bibinfo {author}
  {\bibfnamefont {G.}~\bibnamefont {Bian}}, \bibinfo {author} {\bibfnamefont
  {C.}~\bibnamefont {Zhang}}, \bibinfo {author} {\bibfnamefont
  {R.}~\bibnamefont {Sankar}}, \bibinfo {author} {\bibfnamefont
  {G.}~\bibnamefont {Chang}}, \bibinfo {author} {\bibfnamefont
  {Z.}~\bibnamefont {Yuan}}, \bibinfo {author} {\bibfnamefont {C.-C.}\
  \bibnamefont {Lee}},  \emph {et~al.},\ }\href@noop {} {\bibfield  {journal}
  {\bibinfo  {journal} {Science}\ }\textbf {\bibinfo {volume} {349}},\ \bibinfo
  {pages} {613} (\bibinfo {year} {2015})}\BibitemShut {NoStop}%
\bibitem [{\citenamefont {Lv}\ \emph {et~al.}(2015)\citenamefont {Lv},
  \citenamefont {Weng}, \citenamefont {Fu}, \citenamefont {Wang}, \citenamefont
  {Miao}, \citenamefont {Ma}, \citenamefont {Richard}, \citenamefont {Huang},
  \citenamefont {Zhao}, \citenamefont {Chen}, \citenamefont {Fang},
  \citenamefont {Dai}, \citenamefont {Qian},\ and\ \citenamefont
  {Ding}}]{Lv2015}%
  \BibitemOpen
  \bibfield  {author} {\bibinfo {author} {\bibfnamefont {B.~Q.}\ \bibnamefont
  {Lv}}, \bibinfo {author} {\bibfnamefont {H.~M.}\ \bibnamefont {Weng}},
  \bibinfo {author} {\bibfnamefont {B.~B.}\ \bibnamefont {Fu}}, \bibinfo
  {author} {\bibfnamefont {X.~P.}\ \bibnamefont {Wang}}, \bibinfo {author}
  {\bibfnamefont {H.}~\bibnamefont {Miao}}, \bibinfo {author} {\bibfnamefont
  {J.}~\bibnamefont {Ma}}, \bibinfo {author} {\bibfnamefont {P.}~\bibnamefont
  {Richard}}, \bibinfo {author} {\bibfnamefont {X.~C.}\ \bibnamefont {Huang}},
  \bibinfo {author} {\bibfnamefont {L.~X.}\ \bibnamefont {Zhao}}, \bibinfo
  {author} {\bibfnamefont {G.~F.}\ \bibnamefont {Chen}}, \bibinfo {author}
  {\bibfnamefont {Z.}~\bibnamefont {Fang}}, \bibinfo {author} {\bibfnamefont
  {X.}~\bibnamefont {Dai}}, \bibinfo {author} {\bibfnamefont {T.}~\bibnamefont
  {Qian}}, \ and\ \bibinfo {author} {\bibfnamefont {H.}~\bibnamefont {Ding}},\
  }\href {\doibase 10.1103/PhysRevX.5.031013} {\bibfield  {journal} {\bibinfo
  {journal} {Phys. Rev. X}\ }\textbf {\bibinfo {volume} {5}},\ \bibinfo {pages}
  {031013} (\bibinfo {year} {2015})}\BibitemShut {NoStop}%
\bibitem [{\citenamefont {Nielsen}\ and\ \citenamefont
  {Ninomiya}(1981)}]{Nielsen1981}%
  \BibitemOpen
  \bibfield  {author} {\bibinfo {author} {\bibfnamefont {H.~B.}\ \bibnamefont
  {Nielsen}}\ and\ \bibinfo {author} {\bibfnamefont {M.}~\bibnamefont
  {Ninomiya}},\ }\href@noop {} {\bibfield  {journal} {\bibinfo  {journal}
  {Physics Letters B}\ }\textbf {\bibinfo {volume} {105}},\ \bibinfo {pages}
  {219} (\bibinfo {year} {1981})}\BibitemShut {NoStop}%
\bibitem [{\citenamefont {Son}\ and\ \citenamefont {Spivak}(2013)}]{Son2013}%
  \BibitemOpen
  \bibfield  {author} {\bibinfo {author} {\bibfnamefont {D.~T.}\ \bibnamefont
  {Son}}\ and\ \bibinfo {author} {\bibfnamefont {B.~Z.}\ \bibnamefont
  {Spivak}},\ }\href {\doibase 10.1103/PhysRevB.88.104412} {\bibfield
  {journal} {\bibinfo  {journal} {Phys. Rev. B}\ }\textbf {\bibinfo {volume}
  {88}},\ \bibinfo {pages} {104412} (\bibinfo {year} {2013})}\BibitemShut
  {NoStop}%
\bibitem [{\citenamefont {Yang}\ \emph {et~al.}(2011)\citenamefont {Yang},
  \citenamefont {Lu},\ and\ \citenamefont {Ran}}]{Yang2011}%
  \BibitemOpen
  \bibfield  {author} {\bibinfo {author} {\bibfnamefont {K.-Y.}\ \bibnamefont
  {Yang}}, \bibinfo {author} {\bibfnamefont {Y.-M.}\ \bibnamefont {Lu}}, \ and\
  \bibinfo {author} {\bibfnamefont {Y.}~\bibnamefont {Ran}},\ }\href {\doibase
  10.1103/PhysRevB.84.075129} {\bibfield  {journal} {\bibinfo  {journal} {Phys.
  Rev. B}\ }\textbf {\bibinfo {volume} {84}},\ \bibinfo {pages} {075129}
  (\bibinfo {year} {2011})}\BibitemShut {NoStop}%
\bibitem [{\citenamefont {Wan}\ \emph {et~al.}(2011)\citenamefont {Wan},
  \citenamefont {Turner}, \citenamefont {Vishwanath},\ and\ \citenamefont
  {Savrasov}}]{Wan2011}%
  \BibitemOpen
  \bibfield  {author} {\bibinfo {author} {\bibfnamefont {X.}~\bibnamefont
  {Wan}}, \bibinfo {author} {\bibfnamefont {A.~M.}\ \bibnamefont {Turner}},
  \bibinfo {author} {\bibfnamefont {A.}~\bibnamefont {Vishwanath}}, \ and\
  \bibinfo {author} {\bibfnamefont {S.~Y.}\ \bibnamefont {Savrasov}},\ }\href
  {\doibase 10.1103/PhysRevB.83.205101} {\bibfield  {journal} {\bibinfo
  {journal} {Phys. Rev. B}\ }\textbf {\bibinfo {volume} {83}},\ \bibinfo
  {pages} {205101} (\bibinfo {year} {2011})}\BibitemShut {NoStop}%
\bibitem [{\citenamefont {{Burkov}}\ \emph {et~al.}(2011)\citenamefont
  {{Burkov}}, \citenamefont {{Hook}},\ and\ \citenamefont
  {{Balents}}}]{Burkov2011}%
  \BibitemOpen
  \bibfield  {author} {\bibinfo {author} {\bibfnamefont {A.~A.}\ \bibnamefont
  {{Burkov}}}, \bibinfo {author} {\bibfnamefont {M.~D.}\ \bibnamefont
  {{Hook}}}, \ and\ \bibinfo {author} {\bibfnamefont {L.}~\bibnamefont
  {{Balents}}},\ }\href {\doibase 10.1103/PhysRevB.84.235126} {\bibfield
  {journal} {\bibinfo  {journal} {\prb}\ }\textbf {\bibinfo {volume} {84}},\
  \bibinfo {eid} {235126} (\bibinfo {year} {2011})}\BibitemShut {NoStop}%
\bibitem [{\citenamefont {Kim}\ \emph {et~al.}(2015)\citenamefont {Kim},
  \citenamefont {Wieder}, \citenamefont {Kane},\ and\ \citenamefont
  {Rappe}}]{Kim2015}%
  \BibitemOpen
  \bibfield  {author} {\bibinfo {author} {\bibfnamefont {Y.}~\bibnamefont
  {Kim}}, \bibinfo {author} {\bibfnamefont {B.~J.}\ \bibnamefont {Wieder}},
  \bibinfo {author} {\bibfnamefont {C.~L.}\ \bibnamefont {Kane}}, \ and\
  \bibinfo {author} {\bibfnamefont {A.~M.}\ \bibnamefont {Rappe}},\ }\href
  {\doibase 10.1103/PhysRevLett.115.036806} {\bibfield  {journal} {\bibinfo
  {journal} {Phys. Rev. Lett.}\ }\textbf {\bibinfo {volume} {115}},\ \bibinfo
  {pages} {036806} (\bibinfo {year} {2015})}\BibitemShut {NoStop}%
\bibitem [{\citenamefont {Liang}\ \emph {et~al.}(2016)\citenamefont {Liang},
  \citenamefont {Zhou}, \citenamefont {Yu}, \citenamefont {Wang},\ and\
  \citenamefont {Weng}}]{QiFeng2016}%
  \BibitemOpen
  \bibfield  {author} {\bibinfo {author} {\bibfnamefont {Q.-F.}\ \bibnamefont
  {Liang}}, \bibinfo {author} {\bibfnamefont {J.}~\bibnamefont {Zhou}},
  \bibinfo {author} {\bibfnamefont {R.}~\bibnamefont {Yu}}, \bibinfo {author}
  {\bibfnamefont {Z.}~\bibnamefont {Wang}}, \ and\ \bibinfo {author}
  {\bibfnamefont {H.}~\bibnamefont {Weng}},\ }\href {\doibase
  10.1103/PhysRevB.93.085427} {\bibfield  {journal} {\bibinfo  {journal} {Phys.
  Rev. B}\ }\textbf {\bibinfo {volume} {93}},\ \bibinfo {pages} {085427}
  (\bibinfo {year} {2016})}\BibitemShut {NoStop}%
\bibitem [{\citenamefont {Zhong}\ \emph {et~al.}(2016)\citenamefont {Zhong},
  \citenamefont {Chen}, \citenamefont {Xie}, \citenamefont {Yang},
  \citenamefont {Cohen},\ and\ \citenamefont {Zhang}}]{Zhong2016}%
  \BibitemOpen
  \bibfield  {author} {\bibinfo {author} {\bibfnamefont {C.}~\bibnamefont
  {Zhong}}, \bibinfo {author} {\bibfnamefont {Y.}~\bibnamefont {Chen}},
  \bibinfo {author} {\bibfnamefont {Y.}~\bibnamefont {Xie}}, \bibinfo {author}
  {\bibfnamefont {S.~A.}\ \bibnamefont {Yang}}, \bibinfo {author}
  {\bibfnamefont {M.~L.}\ \bibnamefont {Cohen}}, \ and\ \bibinfo {author}
  {\bibfnamefont {S.}~\bibnamefont {Zhang}},\ }\href@noop {} {\bibfield
  {journal} {\bibinfo  {journal} {Nanoscale}\ }\textbf {\bibinfo {volume}
  {8}},\ \bibinfo {pages} {7232} (\bibinfo {year} {2016})}\BibitemShut
  {NoStop}%
\bibitem [{\citenamefont {Agterberg}\ \emph {et~al.}(2017)\citenamefont
  {Agterberg}, \citenamefont {Brydon},\ and\ \citenamefont
  {Timm}}]{Agterberg2017}%
  \BibitemOpen
  \bibfield  {author} {\bibinfo {author} {\bibfnamefont {D.~F.}\ \bibnamefont
  {Agterberg}}, \bibinfo {author} {\bibfnamefont {P.~M.~R.}\ \bibnamefont
  {Brydon}}, \ and\ \bibinfo {author} {\bibfnamefont {C.}~\bibnamefont
  {Timm}},\ }\href {\doibase 10.1103/PhysRevLett.118.127001} {\bibfield
  {journal} {\bibinfo  {journal} {Phys. Rev. Lett.}\ }\textbf {\bibinfo
  {volume} {118}},\ \bibinfo {pages} {127001} (\bibinfo {year}
  {2017})}\BibitemShut {NoStop}%
\bibitem [{\citenamefont {{Timm}}\ \emph {et~al.}()\citenamefont {{Timm}},
  \citenamefont {{Schnyder}}, \citenamefont {{Agterberg}},\ and\ \citenamefont
  {{Brydon}}}]{Timm2017}%
  \BibitemOpen
  \bibfield  {author} {\bibinfo {author} {\bibfnamefont {C.}~\bibnamefont
  {{Timm}}}, \bibinfo {author} {\bibfnamefont {A.~P.}\ \bibnamefont
  {{Schnyder}}}, \bibinfo {author} {\bibfnamefont {D.~F.}\ \bibnamefont
  {{Agterberg}}}, \ and\ \bibinfo {author} {\bibfnamefont {P.~M.~R.}\
  \bibnamefont {{Brydon}}},\ }\href@noop {} {\ }\Eprint
  {http://arxiv.org/abs/1707.02739} {arXiv:1707.02739} \BibitemShut {NoStop}%
\bibitem [{\citenamefont {Bzdu\ifmmode~\check{s}\else \v{s}\fi{}ek}\ and\
  \citenamefont {Sigrist}(2017)}]{Bzdusek2017}%
  \BibitemOpen
  \bibfield  {author} {\bibinfo {author} {\bibfnamefont {T.}~\bibnamefont
  {Bzdu\ifmmode~\check{s}\else \v{s}\fi{}ek}}\ and\ \bibinfo {author}
  {\bibfnamefont {M.}~\bibnamefont {Sigrist}},\ }\href {\doibase
  10.1103/PhysRevB.96.155105} {\bibfield  {journal} {\bibinfo  {journal} {Phys.
  Rev. B}\ }\textbf {\bibinfo {volume} {96}},\ \bibinfo {pages} {155105}
  (\bibinfo {year} {2017})}\BibitemShut {NoStop}%
\bibitem [{Note1()}]{Note1}%
  \BibitemOpen
  \bibinfo {note} {One can imagine a scenario where a closed nodal surface is
  partially gapped to a nodal surface with boundaries. This case is not
  discussed in this paper.}\BibitemShut {Stop}%
\bibitem [{\citenamefont {Fang}\ \emph {et~al.}(2015)\citenamefont {Fang},
  \citenamefont {Chen}, \citenamefont {Kee},\ and\ \citenamefont
  {Fu}}]{Fang2015}%
  \BibitemOpen
  \bibfield  {author} {\bibinfo {author} {\bibfnamefont {C.}~\bibnamefont
  {Fang}}, \bibinfo {author} {\bibfnamefont {Y.}~\bibnamefont {Chen}}, \bibinfo
  {author} {\bibfnamefont {H.-Y.}\ \bibnamefont {Kee}}, \ and\ \bibinfo
  {author} {\bibfnamefont {L.}~\bibnamefont {Fu}},\ }\href {\doibase
  10.1103/PhysRevB.92.081201} {\bibfield  {journal} {\bibinfo  {journal} {Phys.
  Rev. B}\ }\textbf {\bibinfo {volume} {92}},\ \bibinfo {pages} {081201}
  (\bibinfo {year} {2015})}\BibitemShut {NoStop}%
\bibitem [{\citenamefont {{Fang}}\ \emph {et~al.}(2016)\citenamefont {{Fang}},
  \citenamefont {{Weng}}, \citenamefont {{Dai}},\ and\ \citenamefont
  {{Fang}}}]{Fang2016}%
  \BibitemOpen
  \bibfield  {author} {\bibinfo {author} {\bibfnamefont {C.}~\bibnamefont
  {{Fang}}}, \bibinfo {author} {\bibfnamefont {H.}~\bibnamefont {{Weng}}},
  \bibinfo {author} {\bibfnamefont {X.}~\bibnamefont {{Dai}}}, \ and\ \bibinfo
  {author} {\bibfnamefont {Z.}~\bibnamefont {{Fang}}},\ }\href {\doibase
  10.1088/1674-1056/25/11/117106} {\bibfield  {journal} {\bibinfo  {journal}
  {Chin Phys B}\ }\textbf {\bibinfo {volume} {25}},\ \bibinfo {eid} {117106}
  (\bibinfo {year} {2016})}\BibitemShut {NoStop}%
\bibitem [{\citenamefont {Ho\ifmmode~\check{r}\else
  \v{r}\fi{}ava}(2005)}]{horava2005}%
  \BibitemOpen
  \bibfield  {author} {\bibinfo {author} {\bibfnamefont {P.}~\bibnamefont
  {Ho\ifmmode~\check{r}\else \v{r}\fi{}ava}},\ }\href {\doibase
  10.1103/PhysRevLett.95.016405} {\bibfield  {journal} {\bibinfo  {journal}
  {Phys. Rev. Lett.}\ }\textbf {\bibinfo {volume} {95}},\ \bibinfo {pages}
  {016405} (\bibinfo {year} {2005})}\BibitemShut {NoStop}%
\bibitem [{\citenamefont {Zhao}\ and\ \citenamefont {Wang}(2013)}]{Zhao2013}%
  \BibitemOpen
  \bibfield  {author} {\bibinfo {author} {\bibfnamefont {Y.~X.}\ \bibnamefont
  {Zhao}}\ and\ \bibinfo {author} {\bibfnamefont {Z.~D.}\ \bibnamefont
  {Wang}},\ }\href {\doibase 10.1103/PhysRevLett.110.240404} {\bibfield
  {journal} {\bibinfo  {journal} {Phys. Rev. Lett.}\ }\textbf {\bibinfo
  {volume} {110}},\ \bibinfo {pages} {240404} (\bibinfo {year}
  {2013})}\BibitemShut {NoStop}%
\bibitem [{\citenamefont {Matsuura}\ \emph {et~al.}(2013)\citenamefont
  {Matsuura}, \citenamefont {Chang}, \citenamefont {Schnyder},\ and\
  \citenamefont {Ryu}}]{Matsuura2013}%
  \BibitemOpen
  \bibfield  {author} {\bibinfo {author} {\bibfnamefont {S.}~\bibnamefont
  {Matsuura}}, \bibinfo {author} {\bibfnamefont {P.-Y.}\ \bibnamefont {Chang}},
  \bibinfo {author} {\bibfnamefont {A.~P.}\ \bibnamefont {Schnyder}}, \ and\
  \bibinfo {author} {\bibfnamefont {S.}~\bibnamefont {Ryu}},\ }\href@noop {}
  {\bibfield  {journal} {\bibinfo  {journal} {New J. Phys.}\ }\textbf {\bibinfo
  {volume} {15}},\ \bibinfo {pages} {065001} (\bibinfo {year}
  {2013})}\BibitemShut {NoStop}%
\bibitem [{Note2()}]{Note2}%
  \BibitemOpen
  \bibinfo {note} {In high energy physics this is known as the axial
  symmetry.}\BibitemShut {Stop}%
\bibitem [{\citenamefont {Fukushima}\ \emph {et~al.}(2008)\citenamefont
  {Fukushima}, \citenamefont {Kharzeev},\ and\ \citenamefont
  {Warringa}}]{Fukushima2008}%
  \BibitemOpen
  \bibfield  {author} {\bibinfo {author} {\bibfnamefont {K.}~\bibnamefont
  {Fukushima}}, \bibinfo {author} {\bibfnamefont {D.~E.}\ \bibnamefont
  {Kharzeev}}, \ and\ \bibinfo {author} {\bibfnamefont {H.~J.}\ \bibnamefont
  {Warringa}},\ }\href {\doibase 10.1103/PhysRevD.78.074033} {\bibfield
  {journal} {\bibinfo  {journal} {Phys.Rev. D}\ }\textbf {\bibinfo {volume}
  {78}},\ \bibinfo {pages} {74033} (\bibinfo {year} {2008})}\BibitemShut
  {NoStop}%
\bibitem [{\citenamefont {{Witten}}(2016)}]{Witten2016}%
  \BibitemOpen
  \bibfield  {author} {\bibinfo {author} {\bibfnamefont {E.}~\bibnamefont
  {{Witten}}},\ }\href {\doibase 10.1393/ncr/i2016-10125-3} {\bibfield
  {journal} {\bibinfo  {journal} {Nuovo Cimento Rivista Serie}\ }\textbf
  {\bibinfo {volume} {39}},\ \bibinfo {pages} {313} (\bibinfo {year} {2016})},\
  \Eprint {http://arxiv.org/abs/1510.07698} {arXiv:1510.07698} \BibitemShut
  {NoStop}%
\bibitem [{\citenamefont {McCormick}\ \emph {et~al.}(2017)\citenamefont
  {McCormick}, \citenamefont {Kimchi},\ and\ \citenamefont
  {Trivedi}}]{McCormick2017}%
  \BibitemOpen
  \bibfield  {author} {\bibinfo {author} {\bibfnamefont {T.~M.}\ \bibnamefont
  {McCormick}}, \bibinfo {author} {\bibfnamefont {I.}~\bibnamefont {Kimchi}}, \
  and\ \bibinfo {author} {\bibfnamefont {N.}~\bibnamefont {Trivedi}},\ }\href
  {\doibase 10.1103/PhysRevB.95.075133} {\bibfield  {journal} {\bibinfo
  {journal} {Phys. Rev. B}\ }\textbf {\bibinfo {volume} {95}},\ \bibinfo
  {pages} {075133} (\bibinfo {year} {2017})}\BibitemShut {NoStop}%
\bibitem [{Note3()}]{Note3}%
  \BibitemOpen
  \bibinfo {note} {\label {fn1}. In addition to the $PT$ symmetry, the nodal
  tori models \protect \cref {mod2} and \protect \cref {mod4} are also
  invariant a unitary mirror ($z\to -z$) symmetry, but we do not discuss its
  consequences in this paper.}\BibitemShut {Stop}%
\bibitem [{Note4()}]{Note4}%
  \BibitemOpen
  \bibinfo {note} {If we break $PT$ symmetry, the block Berry phase can be
  changed continuously.}\BibitemShut {Stop}%
\bibitem [{\citenamefont {Rui}\ \emph {et~al.}(2017)\citenamefont {Rui},
  \citenamefont {Zhao},\ and\ \citenamefont {Schnyder}}]{rui2017}%
  \BibitemOpen
  \bibfield  {author} {\bibinfo {author} {\bibfnamefont {W.}~\bibnamefont
  {Rui}}, \bibinfo {author} {\bibfnamefont {Y.}~\bibnamefont {Zhao}}, \ and\
  \bibinfo {author} {\bibfnamefont {A.~P.}\ \bibnamefont {Schnyder}},\
  }\href@noop {} {\bibfield  {journal} {\bibinfo  {journal} {arXiv:1703.05958}\
  } (\bibinfo {year} {2017})}\BibitemShut {NoStop}%
\bibitem [{\citenamefont {Wang}\ and\ \citenamefont
  {Nandkishore}(2017)}]{Wang2017}%
  \BibitemOpen
  \bibfield  {author} {\bibinfo {author} {\bibfnamefont {Y.}~\bibnamefont
  {Wang}}\ and\ \bibinfo {author} {\bibfnamefont {R.~M.}\ \bibnamefont
  {Nandkishore}},\ }\href {\doibase 10.1103/PhysRevB.95.060506} {\bibfield
  {journal} {\bibinfo  {journal} {Phys. Rev. B}\ }\textbf {\bibinfo {volume}
  {95}},\ \bibinfo {pages} {060506} (\bibinfo {year} {2017})}\BibitemShut
  {NoStop}%
\bibitem [{\citenamefont {Kobayashi}\ \emph {et~al.}(2014)\citenamefont
  {Kobayashi}, \citenamefont {Shiozaki}, \citenamefont {Tanaka},\ and\
  \citenamefont {Sato}}]{Kobayashi2014}%
  \BibitemOpen
  \bibfield  {author} {\bibinfo {author} {\bibfnamefont {S.}~\bibnamefont
  {Kobayashi}}, \bibinfo {author} {\bibfnamefont {K.}~\bibnamefont {Shiozaki}},
  \bibinfo {author} {\bibfnamefont {Y.}~\bibnamefont {Tanaka}}, \ and\ \bibinfo
  {author} {\bibfnamefont {M.}~\bibnamefont {Sato}},\ }\href {\doibase
  10.1103/PhysRevB.90.024516} {\bibfield  {journal} {\bibinfo  {journal} {Phys.
  Rev. B}\ }\textbf {\bibinfo {volume} {90}},\ \bibinfo {pages} {024516}
  (\bibinfo {year} {2014})}\BibitemShut {NoStop}%
\bibitem [{\citenamefont {Zhao}\ \emph {et~al.}(2016)\citenamefont {Zhao},
  \citenamefont {Schnyder},\ and\ \citenamefont {Wang}}]{Zhao2016}%
  \BibitemOpen
  \bibfield  {author} {\bibinfo {author} {\bibfnamefont {Y.~X.}\ \bibnamefont
  {Zhao}}, \bibinfo {author} {\bibfnamefont {A.~P.}\ \bibnamefont {Schnyder}},
  \ and\ \bibinfo {author} {\bibfnamefont {Z.~D.}\ \bibnamefont {Wang}},\
  }\href {\doibase 10.1103/PhysRevLett.116.156402} {\bibfield  {journal}
  {\bibinfo  {journal} {Phys. Rev. Lett.}\ }\textbf {\bibinfo {volume} {116}},\
  \bibinfo {pages} {156402} (\bibinfo {year} {2016})}\BibitemShut {NoStop}%
\bibitem [{Note5()}]{Note5}%
  \BibitemOpen
  \bibinfo {note} {Since $\pi _1[SO(n)]=\protect \mathbb {Z}_2$ for $n>2$, the
  winding number is only $\protect \mathbb {Z}_2$-valued for models with more
  than four bands \cite {Bzdusek2017}.}\BibitemShut {Stop}%
\end{thebibliography}%

\end{document}